\documentclass[sigconf,table,xcdraw,nonacm]{acmart}
\AtBeginDocument{%
  }

\usepackage[nameinlink]{cleveref}
\usepackage[most]{tcolorbox}

\usepackage{enumitem}
\usepackage{tikz}
\usetikzlibrary{shapes.geometric}
\usepackage[commandnameprefix=always, final]{changes}
\newcommand{\chibullet}{%
  \tikz[baseline=-0.3ex]{
    \draw[fill, rounded corners=0.4pt]
      (0,0) rectangle (0.75ex,0.75ex);
  }%
}

\newlist{chilist}{itemize}{1}
\setlist[chilist]{
  label=\chibullet,
  labelsep=0.6em,
  leftmargin=1.6em,
  topsep=0.35\baselineskip,
  itemsep=0.15\baselineskip,
  parsep=0pt
}

\newtcolorbox{standout}[2][]{%
  enhanced,
  breakable,
  colback=blue!2!white,
  colframe=blue!70!black,
  colbacktitle=blue!12!white,
  coltitle=black,
  fonttitle=\bfseries,
  title={#2},
  attach boxed title to top left={yshift*=-\tcboxedtitleheight/2, xshift=8pt},
  boxed title style={boxrule=0.8pt, sharp corners, arc=2mm},
  left=10pt, right=10pt, top=8pt, bottom=8pt,
  arc=2mm,
  boxrule=0.8pt,
  drop shadow,
  #1 
}

\newtcolorbox[use counter=democounter,
 crefname={demonstration}{demonstrations}]{demonstration}[2][]{%
  enhanced,
  breakable,
  colback=blue!2!white,
  colframe=blue!70!black,
  colbacktitle=blue!12!white,
  coltitle=black,
  fonttitle=\bfseries,
  title={Demo \thetcbcounter: #2},
  attach boxed title to top left={yshift*=-\tcboxedtitleheight/2, xshift=8pt},
  boxed title style={boxrule=0.8pt, sharp corners, arc=2mm},
  left=10pt, right=10pt, top=8pt, bottom=8pt,
  arc=2mm,
  boxrule=0.8pt,
  drop shadow,
  #1 
}

\crefname{democounter}{Demonstration}{Demonstrations}
\Crefname{democounter}{Demonstration}{Demonstrations}

\begin{document}

\title{Embodied Explainability and Ontological Obstacles}
\subtitle{Why We Struggle to Explain the Answers of Large Language Models (LLMs)}

\author{Marvin Pafla}
\email{mpafla@uwaterloo.ca}
\affiliation{%
  \institution{University of Waterloo}
  \streetaddress{200 University Ave W}
  \city{Waterloo}
  \state{Ontario}
  \country{Canada}
  \postcode{N2L 3G1}
}

\author{Jesse Hoey}
\email{jesse.hoey@uwaterloo.ca}
\affiliation{%
  \institution{University of Waterloo}
  \streetaddress{200 University Ave W}
  \city{Waterloo}
  \state{Ontario}
  \country{Canada}
  \postcode{N2L 3G1}
}

\author{Kate Larson}
\email{kate.larson@uwaterloo.ca}
\affiliation{%
  \institution{University of Waterloo}
  \streetaddress{200 University Ave W}
  \city{Waterloo}
  \state{Ontario}
  \country{Canada}
  \postcode{N2L 3G1}
}

\author{Mark Hancock}
\email{mark.hancock@uwaterloo.ca}
\affiliation{%
  \institution{University of Waterloo}
  \streetaddress{200 University Ave W}
  \city{Waterloo}
  \state{Ontario}
  \country{Canada}
  \postcode{N2L 3G1}
}






\renewcommand{\shortauthors}{Pafla et al.}

\begin{abstract}
    Explainability is often framed as a property of an AI model, with explanations extracted from its internals and shown to users. In this argument paper, we instead provide an embodied account of explainability based on Dourish and enactivist cognition: understanding is created in use as people act on affordances in shared practice. Using demonstrations and conceptual analysis, we reveal ontological obstacles when ``looking inside'' large language models: surrogates import external abstractions that can be mistaken for the model's, and focusing on internal reasoning misses that explainers participate in their own understanding. We discuss these obstacles in XAI practice, arguing that many explanations are misnamed, which skews their purpose and can increase overreliance. Finally, we highlight how embodied explanations reorganize sense-making by making what matters publicly available for action, and argue that explainability claims should be reserved for designs that provide affordances to probe, coordinate, and repair behaviour in situated practice.
\end{abstract}


\maketitle

\section{Introduction}


Large language models (LLMs) increasingly mediate consequential work, and with that comes a renewed demand: they should be \emph{explainable}. Across HCI and XAI, this demand is addressed in a variety of ways---for example by attaching post-hoc artifacts (e.g., feature attributions, natural-language rationales), building simplified surrogates, or ``looking inside'' via mechanistic interpretability. Despite their differences, these approaches share a common hope: that exposing the model's relevant ``reasons'' will help people decide when to trust, contest, or override its outputs.

This view of explainability is increasingly strained in active practice. Empirical studies show that XAI artifacts (e.g., saliency maps) can be persuasive without being helpful, and can even increase overreliance when they accompany incorrect outputs \cite{overreliance2025llms, bansal2021, veri_fok_2023, manipulating2021, si-etal-2024-large, our2024chi, cxaiallyouneed, imperfectxai}. In many chat-style LLM deployments, fluent rationales can mimic understanding while leaving users without reliable ways to probe what would change an output, diagnose failures, or coordinate responsibility \cite{our2024chi, overreliance2025llms, cxaiallyouneed}. This finding creates a practical dilemma: demand for explainability rises, but these artifacts often fail to provide the stable affordances needed for intervention and repair in situated activity \cite{ailaw, lawyer_per, Sokol_Flach_one_expl_does_not_fit, miller2018explanation}. The stakes extend beyond usability: regulation and liability demand lay-understandable explanations and freedom from defects, yet many existing approaches often reveal defects without guaranteeing their absence and leave it unclear when an explanation is ``good enough'' \cite{ailaw, gdpr_expl}.

We argue that this mismatch is not only technical but conceptual. Many XAI methods implicitly treat the explanandum as fixed in advance, assuming that understanding follows from exposing the ``right'' internal structure \cite{doshi-valez_towards_interpret_ml, smith_ooo}. This framing inherits a familiar \emph{internalist} (``Cartesian'') standpoint in which cognition is a matter of internal representations of a mind-independent world \cite{smith_ooo, Gallagher2023}. It also aligns with a code model and theory-of-mind framing of communication: meaning is treated as private content to be encoded and decoded by inferring what is ``in the head'' of the agent \cite{scott2014speaking}.

But lived activity rarely fits this picture. What counts as relevant is not simply ``there'' to be read off and mapped to an internal rationale; it is grounded in practice through roles, tools, timing, and shared routines, and can shift abruptly under breakdown or changing stakes \cite{Dourish2004, kirchhoff2019dc, DeJaegher2007}. We therefore take \emph{embodied explainability} as our starting point: a system is explainable insofar as it offers \emph{affordances} people can use to coordinate, check, and repair what is going on \cite{dourish2001action, Dourish2004}. Such affordances are not fixed properties of an interface; they come into view through what people can actually do with it. In short, understanding is not merely delivered from the inside out; \emph{explainers participate in it}.

On this account, we run into a familiar tension: the very scale that makes LLMs powerful also makes them difficult to make understandable in use. Their performance comes from high-dimensional, distributed structure, while explainability in practice depends on a small set of stable factors people can manipulate, check, and share. Many XAI approaches---including mechanistic interpretability---respond by building \emph{surrogates}: they compress the model into legible objects (saliency, rationales, features, circuits, graphs) that can be made intelligible. These surrogates can be valuable for analysis and debugging, but they introduce what we call \textit{ontological obstacles} when treated as explanations for situated use: to be usable, one must impose abstractions that can look like ``the model's concepts'' but are often supplied by the explainer. Separately, the fixation on ``looking inside'' can miss \emph{where the action is} \cite{dourish2001action}---the embodied, social work through which people coordinate, test, and repair understanding at the point of use.

If ``looking inside'' yields partial, abstracted structure that rarely functions as an everyday interactional resource, then explainability may not be settled by internal reasons alone. We therefore shift from internal reasons to how explanations become meaningful in public interaction, drawing on a pragmatic account of communication in which agents make intentions evident and others interpret that evidence against what is relevant in context \cite{scott2014speaking, grice1982meaning}. This framing encourages building systems that can join the public organization of the activity---and keeping that activity alive---rather than scaling context and adding explanations after the fact. Separate from HCI's calls for accountability and transparency, we thus make an ontological point that connects understanding with the existence of interactional phenomena like conversations or queues.

\begin{sloppypar}
We develop the argument through conceptual analysis and demonstration-based reasoning: short scenarios that show what an explanation must accomplish to be usable in practice. In \Cref{sec:embodied}, the hammer (\Cref{demo:i-can}) motivates an embodied view of objects and understanding. In \Cref{sec:ex_is_em}, the clinic contrast (\Cref{demo:ontological-problems}) highlights the importance of embodiment in explainability. In \Cref{sec:oo}, we analyze ``looking inside'' LLMs as surrogate-building and argue that it runs into ontological obstacles. In \Cref{sec:xai_llm}, we situate LLM explainability within XAI and synthesize empirical findings. In \Cref{sec:public_explanations}, the queue breakdown (\Cref{demo:queue}) shifts the focus from internal reasons to public repair of shared sense-making. Finally, \Cref{sec:discussion} frames explainability as a commitment to situated participation and accountable coupling, not an extra output channel layered onto general-purpose chat.
\end{sloppypar}

This paper makes four contributions. First, we offer a unified account of \emph{embodied explainability} that links affordances, the constitution of objects and context, and public meaning. Second, we reveal \emph{ontological obstacles} when ``explaining'' LLMs via mechanistic interpretability and a focus on the model's internals. Third, we frame the purpose of explanations in the larger context of XAI. Fourth, we explore \emph{public explanations}: ostensive, norm-setting interventions that stabilize what is happening and what follows next by installing shared affordances for coordination and repair.

\section{Embodied Cognition}
\label{sec:embodied}

This section introduces the embodied-cognition lens underlying our account of explainability. We show how environments appear to agents in terms of action possibilities, using the hammer to highlight that what a thing is is inseparable from skilled use. We then argue that objects and context are stabilized through ongoing activity rather than given in advance. Because fixed objects and categories can miss what matters in practice, this section sets up our later argument that explanation should be located in public interaction, and clarifies why XAI approaches that ``look inside'' run into ontological obstacles.

\subsection{Affordances}

We argue that the different ways embodied agents can interact with the real world constitute how they will perceive and act in it. While many researchers in HCI are likely familiar with the concept of affordances, it is typically applied to interactive devices or techniques (e.g., \citet{norman1988psychology}), but our emphasis in this work is on how our actions bring forth the world we live in. We thus make an ontological argument about what the world is made out of---what objects, phenomena, and things \textit{exist}. To understand this idea, it helps to question the existence of objects as completely independent of the mind or body---a line of thought we invite with the first demonstration.

\begin{demonstration}[breakable=false, label={demo:i-can}]{Affordances}
Consider a basic object like a hammer. In fact, the word ``hammer'' might already trigger a plethora of associations and meanings in the reader such as its affordance, i.e., the activity this object is designed \textit{for}, its principal components and how it feels to use such a hammer. But let's imagine this hammer is right in front of us when we try to describe it. How would we do this? We might try to describe its components such as its wooden handle or its shining metallic head. In doing so, one might realize that such descriptions can feel arbitrary; for example, even if it had a handle made of hardened plastic, we would still call the object a hammer. Furthermore, we might leave out certain details of the hammer such as the scratch along its head that was caused when missing the last nail necessary to finish the wooden deck last spring.
\Description{A demonstration discussing the affordances of a hammer.}
\end{demonstration}

\Cref{demo:i-can} shows that simply arranging matter in a certain way is not enough for it to show up \emph{as} a hammer: its ``thingness'' arrives through embodied \emph{sense-making}, as pre-reflective, skilful possibilities of use (grasping, swinging, hitting nails) become immediately salient. Husserl describes these as ``I can'' relations, and Merleau-Ponty links them to a \emph{body schema}---embodied know-how that organizes what actions are available \cite{husserl1989ideas, merleau2012phenomenology}. On this view, what counts as \emph{the} object is co-constituted in agent--environment coupling: as skills and practices change, so too do the object's boundaries, relevant parts, and even its type. In fluent tool use, the object becomes \emph{ready-to-hand} (embodied in action and barely noticed), whereas in breakdown (e.g., when the tool fails) it becomes \emph{present-at-hand}, turning into a set of parts and explicit thoughts \emph{about} the tool \cite{heidegger1962being, smith_ooo}; this distinction has also informed HCI work on tool embodiment in VR \cite{alzayat2019toolembodiment}.

This analysis motivates the notion of \emph{affordances}: following \citet{gibson2014ecological}, environments offer agents (human or AI) possibilities for action that are \emph{directly perceivable} in situated engagement. In HCI, Gibson's affordances entered through design-oriented questions about what technologies are \emph{for}: how artifacts, interfaces, and devices can make their action possibilities easy to see and learn. Early work confronted mismatches between what objects allow and what users perceive, prompting Norman's affordance/signifier distinction to explain false or hidden cues \cite{norman1988psychology}. Gaver urged making affordances detectable---often in sequences---and cautioned about timing and context \cite{gaver1991}. As digital systems became popular, usage fragmented, motivating calls to separate possibilities from cues \cite{MH-gi2000} and, with the social turn, to foreground ``cultural affordances'' shaped by norms and shared practice \cite{Ramstead2016-zd}. Recent proposals add a computational lens, modelling how users infer likely affordances under uncertainty and cost \cite{redefiningaffordance}.

While the body anchors affordances by tightly coupling perception and action (as seen in motor resonance and action–language effects \cite{Goldman2009Vignemont, Goldman2012, Pulvermüller2005}, in systematic shifts in judgment and perceptual experience with bodily state \cite{danziger2011, Proffitt1995, VARGA201777, Zelano12448}, and in the different affordance structures disclosed by different bodies \cite{Goyal2023, Gallagher2023}), this dependence does not mean cognition is sealed within a single individual. When people coordinate with tools, notations, and one another, cognition becomes \emph{distributed} \cite{hutchins1995, hutchins2000distributed}: parts of the cognitive work are offloaded into the environment and social systems. Over developmental, cultural, and evolutionary time, communities build \emph{cognitive niches}---stable arrangements of artifacts, practices, and norms that shape what stands out as relevant and what counts as a sensible next move \cite{Sterelny2010}. 

From this perspective, \textit{understanding} is a stabilized, revisable practical grip that sustains and repairs patterns of sense-making across contexts---including a sensorimotor ``if I do this, that will happen'' know-how \cite{ORegan2001-fz}, often scaffolded by shared tools, routines, and coordination with others \cite{hutchins1995, Dourish2004}. \citeauthor{merleau2012phenomenology}'s \emph{intentional arc}---the sedimented history that binds body, situation, and practice---attunes individual sensorimotor repertoires (i.e., understanding) to niches and public routines \cite{merleau2012phenomenology}. The result is \emph{common ground}: shared expectations, norms, and cues that let multiple people see the same affordances in a situation, so that in \Cref{demo:i-can} the object shows up to us as a hammer because we have hammered, passed hammers, watched others drive nails, and learned the norms of their use.

\subsection{Objects and Context Are Brought Forth in Action}
\label{sec:cons}

This section stresses the ontological implications of affordances: objects and context are brought forth in situated interaction, so how a thing is used or taken up determines how it is perceived as something. Because embodied action is flexible, the variety of objects that can be enacted in practice usually exceeds what any fixed labelled dataset can discretely map, with important implications for XAI approaches that explain in terms of pre-defined categories.

Despite this open-endedness of embodied action, AI often defaults to an internalist picture in which a model maps ``affordance representations'' onto a fixed, mind-independent world \cite{smith_ooo, Gallagher2023}. This shows up in ML/robotics pipelines that ``detect'' affordances via ground-truth labels (e.g., assigning ``cutting'' to a blade) \cite{10171410, 7523298}. But treating affordances as internal representations immediately runs into the hard problem of content: what, exactly, fixes their representational content, and how that content stays aligned with a changing world \cite{Akins1996-AKIOSS-2, Hutto2015-HUTTNO-4}.

Enactive and embodied approaches start from a different place. They take the basic unit of mind to be the active coupling of brain, body, and environment. On this view, objects and reasons do not pre-exist ``inside the head'' as static representations; they are enacted (brought forth) by the ongoing dynamics of situated action \cite{Thompson2005, Gallagher2023}. A classic example is a baseball fielder chasing a fly ball: they do not need to explicitly compute the ball's trajectory with the help of representations they then update. By maintaining a particular optical relation to the ball and adjusting stride in real time, the agent–environment coupling solves the task online, helping explain why fielders keep their eyes on the ball throughout its flight \cite{clark2015, Fink2009-cf}.

Treating brain–body–environment coupling as integral to cognition raises a familiar objection \textit{where} cognition really is: extended/enactive views allegedly ``confuse cause and constitution'' \cite[p. 499]{Carruthers2015498} since environments might merely influence thinking rather than be part of it \cite{block2005review, adams2001bounds, Kiverstein2023, KirchhoffKiverstein2024}. This objection presumes a standard, snapshot-style (or \textit{synchronic}) model of constitution---like a block of marble constituting Michelangelo's David only insofar as they coincide at a time in space \cite{Wilson2007-WILAPA, wasserman2004}---and then argues that what merely causes cognition cannot constitute it (because a cause must come \textit{before} its effect) \cite{block2005review, adams2001bounds, kirchhoff2019dc}. 

\begin{sloppypar}
Enactivists instead appeal to a temporally extended (or \textit{diachronic}) notion of constitution that fits process-based systems \cite{kirchhoff2019dc}. For example, the wave (also Mexican wave) is constituted by norm-guided coordination unfolding over time across two coupled micro-dynamics (individual stand–sit rhythms of people) and macro-dynamics (the travelling pattern across the stadium), where the emergent pattern in turn constrains local behaviour by setting up expectations about when and how to move \cite{Jaegher2010441, Hutchins2011437, kirchhoff2019dc, KirchhoffKiverstein2024}. The result is a single, self-organizing dynamical system rather than a static sum of parts.
\end{sloppypar}

The wave shows that context is an interactional achievement, not a pre-given background: it is whatever gets recruited into the ongoing coordination that sustains the phenomenon (e.g., adjacency, sightlines, timing norms, bodily rhythm), while most other facts are irrelevant (e.g., what people wear, their personalities, which game they watch). Enactivists treat this kind of norm-guided coordination as a simple case of \emph{participatory sense-making}: as interaction becomes self-sustaining, meaning is enacted in the coupling itself \cite{DeJaegher2007, Gallagher2008Intersub, kirchhoff2019dc}. In this sense, context is relational and inseparable from activity---brought forth by the process it stabilizes---not everything that is true in the stadium. This finding sharpens Dourish's warning \cite{Dourish2004} that context is not just information: stuffing ever more information into a larger context window for LLMs will miss where relevance and meaning are enacted in practice.

\section{Embodied Explainability}
\label{sec:ex_is_em}

Following Dourish \cite{dourish2001action}, we treat explainability as an embodied, interactional achievement: users gain understanding when systems expose stable public affordances they can use to act and reason. Meaning is not delivered by designers but made in situated action---even in ``classical'' screen–keyboard–mouse interaction---because what matters is how a system's categories and resources couple to ongoing practices. Design should therefore avoid a single fixed ontology and instead offer interactional handles: revisitable structures with dependable consequences that support accountable ``what-if'' reasoning \cite{dourish2001action}. This motivates Dourish's critique of ``disappearing'', ubiquitous interfaces: when systems recede from view, users lose traction on the structures they must contest and coordinate around \cite{dourish_epilogue}. Explainability, on this view, is the availability of such public handles in interaction (i.e., affordances), not an internal property to be extracted and displayed.

The focus thus moves from ``explainability'' as a property of the system to meaningful coupling. To exemplify this, \Cref{demo:ontological-problems} contrasts two deployments: one that makes a decision rule available as a public, editable structure, and one that makes a decision available primarily through conversational exchange.

\begin{demonstration}[label={demo:ontological-problems}]{An Explainable Linear Model vs. a Chatbot}

\textbf{Setting.} An urgent-care clinic uses decision support to decide whether to \emph{fast-track} a chest-pain patient for same-day cardiology review.

\textbf{Linear score.} The clinic uses a simple rule:
\[
\begin{aligned}
s(x) ={}& 1.0\,(\text{ECG abnormal}) + 1.0\,(\text{troponin elevated})\\
       &{}+ 0.5\,(\text{age}>65) - 1.0\,(\text{clear non-cardiac cause})
\end{aligned}
\]
(each term is 0 or 1). Fast-track if $s(x)\ge 2$.\\

A nurse, \emph{Priya}, enters: ECG abnormal = 1, troponin elevated = 0 (pending), age${}>65$ = 1, clear non-cardiac cause = 0, so $s(x)=1.5$ (no fast-track). When troponin posts elevated, she flips its field to 1 and $s(x)=2.5$ (fast-track). The physician, \emph{Dana}, asks why; Priya answers: ``ECG abnormal plus troponin elevated.''

\textbf{LLM assistant.} The clinic later adds a chat-based LLM assistant. Staff paste the same information; it replies \texttt{fast-track} with a rationale that cites age and ECG while treating troponin as uncertain.

\begin{quote}
\emph{Dana:} ``What, exactly, would flip this to \texttt{no fast-track}?''\\
\emph{LLM:} ``If troponin comes back normal and the ECG is less concerning, you could avoid fast-tracking; age is still a risk factor.''\\
\emph{Dana:} ``So if I change \emph{only} troponin from elevated to normal, what happens?''\\
\emph{LLM:} ``It depends on the overall picture; given age and abnormal ECG, I would still lean \texttt{fast-track}.''
\end{quote}

When Priya updates troponin from pending to elevated and asks again, the recommendation and rationale shift, but not in a way clinicians can check against a stable set of factors with defined effects.
\end{demonstration}

In the linear case, the model's factorization \textit{is} an affordance: the categories are the rule, and changing one documented fact has a defined effect on the output (troponin $0\!\rightarrow\!1$ adds $+1.0$). That makes counterfactuals actionable (``if troponin had been higher, the decision would have flipped'') and justifications accountable (``ECG abnormal plus troponin elevated''), because staff can inspect, predict, and contest the mapping while they work.

A chat-based LLM can mimic this surface form---it can speak the clinic's language and offer plausible rationales on demand---but the relation between cited factors and the recommendation is not exposed as a small, shared set of commitments that must update in determinate ways under specific edits. The exchange above makes this visible: Dana asks for a crisp \emph{edit-to-effect} relation (``change only troponin''), but the LLM answers with elastic dependencies (``it depends on the overall picture'') rather than a commitment that can be checked. Dana's ``what would change your decision?'' is a request for dependable coupling, not more narrative. The linear score supplies that coupling by design: the effect of changing one field is defined and stable. The LLM, by contrast, produces answers that are locally sensible but not governed by a small, shared factorization that constrains how its reasons must change under specific edits. When Priya updates troponin from ``pending'' to ``elevated,'' the LLM's rationale shifts, but not in a way that reliably\footnote{Note that the LLM can happen to provide this causal dependency while only doing associative token prediction, but the point is exactly that there is no guarantee for that. Likewise, the LLM can sit on top of a model that would by itself be explainable in interaction, but the output of the LLM can obscure a reliable coupling. What the system provides ultimately is \textit{text} that must be interpreted and validated, rather than a public structure that can be acted on directly.} tracks ``this field changed, therefore the decision changed in this particular way.''

Explainability, on this embodied view, is not a matter of fluent text or access to internal traces. It depends on whether a system provides stable, public affordances that users can appropriate to anticipate, coordinate, and justify action. The linear model is explainable here \textit{to} users because its factors coincide with the clinic's working categories and support reliable coupling. By contrast, an LLM may produce plausible rationales on demand, but without a stable, editable factorization that binds edits to determinate effects, those outputs do not function as affordances. This is not a ``linear good, LLM bad,'' argument but an explainability criterion: does the system expose persistent structure users can inspect, modify, and hold accountable? Many LLM deployments as open-ended chat do not, and so fall short of embodied explainability in this local sense.

With this embodied view, we can narrow the scope of our argument: we target settings where LLMs are deployed primarily as open-ended chat interfaces that do not expose persistent affordances and do not bind user edits to a small, checkable set of commitments; in those conditions, ``asking why'' elicits fluent rationales or internal saliency maps (discussed in the next section) but not the dependable coupling required for accountable action. The linear model is locally explainable to users---why this answer in this case---precisely because users can manipulate a compact factorization and anticipate how outputs will change; and when systems add explicit state (e.g., tool traces, simulators, or editable intermediate representations), embodied explainability may be achievable in larger models as well, which we treat as a promising direction rather than a counterexample. This point is also not a rejection of XAI wholesale (\Cref{sec:xai_llm}): many methods work for bounded tasks and expert workflows, where explanations are tied to stable artifacts and intervention points. Our claim is narrower: for everyday use of LLMs to produce long-form answers, we often lack reliable, user-facing handles that bind edits to checkable effects, even as \emph{global} explanations often succeed through the practical couplings of building, probing, and repairing models. 

\section{Ontological Obstacles}
\label{sec:oo}

This section explains why ``looking inside'' large language models often doesn't yield usable explanations. To make a vast model intelligible, we must abstract it into smaller, interpretable objects, and in doing so we inevitably introduce human-made categories and boundaries. We use mechanistic interpretability as a concrete case---both to show what such abstraction can achieve and to surface the ontological obstacles it creates---before turning to the broader pitfalls of treating a model's internal structure as the primary site of explanation.

\subsection{Abstraction in Mechanistic Interpretability}
\label{sec_mec-inter}

We dissect Anthropic's work on the ``biology'' of LLMs \cite{lindsey2025biology} as a case study of mechanistic interpretability and an attempt to abstract complex neural nets into interpretable surrogate models. While this distillation achieves some level of explainability, we argue that this explainability is ultimately propped up by \textit{ontological scaffolds} (e.g., categories, relationships, labels) provided by humans trying to understand the model. 

In Anthropic's companion paper, \citet{ameisen2025circuit} describe a mechanistic-interpretability pipeline that builds interpretable surrogate models meant to mirror an LLM's conceptual processing. The approach assumes an internalist picture in which concepts correspond to features in activations, but these are hard to isolate because they are distributed across ``polysemantic'' neurons \cite{elhage2022superposition, ameisen2025circuit}. Using cross-layer transcoders \cite{dunefsky2024transcoders}, they extract sparse attribution graphs of active features and their interactions, then prune and validate edges via perturbation tests to produce an explorable circuit-level surrogate.

\citet{lindsey2025biology} present compelling illustrations of this internal circuitry---or ``biology''---of LLMs such as Claude 3.5. Most prominently, the authors find evidence for multi-step reasoning: the model deduces, for example, that the state containing Dallas is Texas, and that the capital of Texas is Austin, to satisfy the prompt ``Fact: the capital of the state containing Dallas is''. 
To demonstrate this, the researchers identify ``supernodes''---clusters of features representing a given concept---such as ``capital'', ``say a capital'', and ``relating to the state of Texas''. 
Through successive intervention and interpretation, attribution graphs emerge that are both human-understandable and representative of the model's internal computations. 

Despite these benefits to interpretability, the authors report that their method yields interpretable graphs for only about 35\% of prompts. It also struggles with long prompts and extended internal reasoning chains, and it does not permit counterfactual inspection. Abstracting a highly complex, deep model introduces further constraints: the absence of attention modelling, large error nodes---``dark matter''---not captured by the abstraction, and difficulty representing inactive or inhibitory circuits. 

The ontological point we raise in this section concerns the abstraction of neurons into features. \citet{ameisen2025circuit} note that their method struggles with this abstraction because the transcoder aims to remain faithful to the original LLM, splitting or absorbing features in ways that are uninterpretable to humans. The researchers attempt to address this problem by manually grouping features into ``supernodes,'' which is a labour-intensive process that underscores human participation in what we call ``ontological scaffolding.'' That is, to make sense of attribution graphs at all, humans must interpret them, intervene on them, and formulate meaningful concepts represented by the features. In this way, the researchers engage in a form of ``epistemological foraging'' to learn about the model's internal mechanics. 

\begin{sloppypar}
More generally, LLMs derive their performance from high-dimensional, uninterpretable circuitry \cite{deep_learning}; on the other hand, we aim to compress such circuitry toward linear decision boundaries that are naturally explainable to us. Historically, the appeal of high-dimensional models like neural networks was precisely that they bypassed the brittleness of discrete, knowledge-based AI, operating directly from rich data (e.g., images and text) \textit{without} first committing to---and maintaining---vast, heterogeneous ontological schemes that earlier AI struggled to build \cite{smith2019}. From this perspective, mechanistic interpretability reads as an effort to undo that move into ``latent space,'' which puts a principal ontological obstacle between the scaling performance of large-scale models and their explainability.
\end{sloppypar}

\subsection{The Pitfalls of ``Looking Inside''}

Across ML, cognitive science, and psychology, the reigning Kuhnian paradigm---in which core scientific assumptions are continually reinforced and philosophically enshrined \cite{kuhn1962, slors2012model}---is internalist and representational: cognition is cast as manipulating inner models of a mind-independent world that conveniently yields ``ground-truth'' data \cite{smith_ooo}. This paradigm legitimizes the human provision---and subsequent internalization---of concepts, categories, and relations as an ontological scaffold for AI, as we have seen in the previous section. Furthermore, it frames explainability as a matter of ``\textit{looking inside}'' the system. 

We argue that the project of mechanistic interpretability runs into what Smith \cite{smith_ooo} calls the ontological wall: the limitations faced by one's own parsing of a theoretical situation into objects, properties, relationships, etc. While \citet{smith_ooo} admits that some of this parsing is always necessary in advance, the danger of inscription errors is high, where ontological assumptions are inscribed or imposed onto a system, and then read back off the system ``as if that constituted an independent empirical discovery or theoretical result'' \citep[p. 50]{smith_ooo}.

Thus, for large language models, the familiar strategy of ``looking inside'' runs up against two hard limits. The first is practical: as the previous section showed through Anthropic's ``biology'' perspective, making sense of state-of-the-art models requires heavy abstraction, bespoke tooling, and intensive expert labour, and even then yields only partial, local coverage (e.g., interpretable graphs for roughly 35\% of prompts) \cite{lindsey2025biology, ameisen2025circuit}. As a result, mechanistic interpretability cannot function as a routine, everyday resource for explanation. The second is ontological: as argued in \Cref{sec:embodied}, the objects of explanation are constituted in interaction---they are brought forth in practice, not waiting fully formed inside the model. If explananda are interactional achievements, then focusing design effort on intervening in a model's internal structure---even if it were possible---misses where the very things we are trying to understand are constituted; the internal structure is, in the words of Dourish, \textit{not} ``where the action is.''

These limits undercut Dourish's route to understanding, where meaning arises through embodied coupling to reliable, public affordances. LLM ``explanations'' rarely provide such handles: when ``looking inside'' works, it typically yields abstractions that do not function as interactional resources. For example, saliency maps can increase confirmation bias rather than support understanding when evaluating model answers \cite{our2024chi} and their attributions shift under minor, practice-irrelevant changes and need not track stable causal dependencies \cite{Kindermans2019, NEURIPS2018_294a8ed2, Ghorbani_Abid_Zou_2019, jain-wallace-2019-attention}. As a result, users lack stable action possibilities for linking their practice to what LLMs are doing, leaving everyday sense-making without affordances for embodied coupling.

The implications for explainability are therefore troubling. This case study leaves us with a double bind: we can scale models for performance, but then their internal workings become effectively out of reach as stable resources for everyday coupling; and when we try to recover access through mechanistic interpretability, we obtain abstractions that rarely function as interactional resources and we miss the ontological target of where explanations are located. Without such affordances that connect users' practices to system behaviour, we lose the embodied route to understanding---leaving no clear path to local explainability for LLMs. The next section broadens this diagnosis by situating it within the wider landscape of XAI methods and empirical findings.

\section{Explainable AI and Large Language Models}
\label{sec:xai_llm}

XAI methods can be useful when they bind a model's output to concrete artifacts people can inspect---often by pointing back to aspects of the input, such as influential words or image regions---so people can surface bias, find errors, and coordinate decisions. In LLMs and other deep networks, however, similar artifacts are often hard to associate with stable mechanisms or reliable intervention points, and they can make incorrect outputs feel well-justified, increasing overreliance. Our aim is therefore not to dismiss XAI, but to clarify a boundary: techniques that are valuable for auditing and debugging do not automatically function as user-facing explanations in situated activity. We therefore use this section to ask what explanations are for in practice, situating LLM explainability within the broader landscape of XAI.

XAI can support bias auditing, data summarization, and classification by surfacing feature-importance signals (e.g., saliency maps) that aid individual and group reasoning and help reveal errors \cite{biasDetector, Ronnback2025-sa, theroleofllm, amplyfyinggroup}. For example, explanations improved group calibration in mushroom-edibility classification \cite{amplyfyinggroup}, highlighted clinically meaningful regions in brain-disease diagnosis \cite{brain_xai}, exposed token-level bias via input–output saliency \cite{biasinllms, human-biases-expl-wang-2019}, and helped users navigate input spaces to detect image-classification errors \cite{humanexp2023morrison}.

\begin{sloppypar}
Much of XAI seeks explanations in compressed, human-explainable abstractions---local surrogates (e.g., LIME/SHAP), provenance traces, or induced rules/programs \cite{lipton_2018, doshi-valez_towards_interpret_ml, xai_tax_2020_arrieta, samek2017explainable, ex_deep_2021_samek, lime, shap}. These tools originated in small, structured predictive settings (e.g., linear regression in \citet{manipulating2021} or decision trees in \citet{Sokol_Flach_one_expl_does_not_fit}) and often specialize in classification \cite{vilone_notions}, but they clash with the sources of LLM performance: deep, distributed, high-dimensional representations \cite{deep_learning, transformer} whose internal regularities are hard to stabilize into a small set of user-facing affordances \cite{elhage2022toymodelssuperposition, fooling_lime}. As dimensionality increases, surrogate-style explanations become brittle and harder to render intuitive or faithful \cite{historical-confalonieri-2021, weight_of_evidence_2021, lindsey2025biology, jain-wallace-2019-attention}, as we discussed in \Cref{sec:oo}. In practice, this conflict often leaves feature-level salience as the most tractable option in high-dimensional input-spaces \cite{danilevsky-etal-2020-survey, our2024chi}.
\end{sloppypar}

Mounting evidence shows that such explanations can \emph{increase} overreliance: people sometimes switch from correct to incorrect answers when shown compelling but misleading rationales \cite{overreliance2025llms, bansal2021, veri_fok_2023, manipulating2021, si-etal-2024-large, our2024chi, cxaiallyouneed, imperfectxai}. This risk is not accidental but architectural:  contemporary models are fundamentally associative, and can ``self-evidence'' their outputs, so that even when a system is wrong, its explanation can still look locally supportive (e.g., saliency highlighting tokens consistent with an incorrect answer) \cite{deep_learning, lipton_2018, our2024chi}. Similarly, chain-of-thought can improve task performance, but models can be unfaithful to their own rationales, and fluent narrative explanations may obscure reliability cues \cite{turpin2023language, arcuschin2025chainofthought, overreliance2025llms, cxaiallyouneed}. This is the dilemma of explainable AI: when a system errs, its explanation may still appear plausible, making errors harder---not easier---to detect and amplifying confirmation bias \cite{our2024chi, cogbias2025chi}. 

More broadly, in many deployments and policy discussions, explainability is treated as an AI-to-human warning signal for model errors, yet current systems do not consistently identify and communicate their own failure modes without external feedback. Explanations remain largely one-way: users are shown rationales, but their corrections seldom seldom yield immediate, inspectable changes in what the system will do next within the same work episode, often ignoring explanation's social, two-way character \cite{miller2018explanation, miscommunicating}. Designing for interactional repair---where users can articulate what went wrong and the system can incorporate that feedback in context---better matches the collaborative and communicative settings in which ``explainability'' is typically invoked \cite{explaininyourway, miscommunicating}. Until then, current XAI methods may still fall short of the desiderata we have set out for situated use (e.g., \cite{ailaw, lawyer_per, Sokol_Flach_one_expl_does_not_fit, miller2018explanation}).

\section{Public Explanations}
\label{sec:public_explanations}

Having shown the ontological obstacles of ``looking inside'' and the risks of post-hoc XAI, we now shift from internal reasons to how explanations become meaningful in public interaction. We draw on a pragmatic account of communication in which agents make intentions evident through what they do and say, and others interpret this evidence against what is relevant in context \cite{scott2014speaking, grice1982meaning}. Like the wave in \Cref{sec:cons}, such ostensive behaviour is what stabilizes meaningful phenomena, as illustrated by the next demonstration:

\begin{demonstration}[label={demo:queue}]{A Queue Dissolves Without Public Repair}
Kai takes ticket \#18 at a clinic and sits down. The screen calls \#7, then \#9. Then it calls \#31. A few minutes later: \#27. Then \#44. People start looking at each other. Someone says, ``Is the screen broken?'' Another mutters, ``Why did I even take a number?'' A couple of patients walk up to the desk to argue their case; others hover near the entrance so they won't be missed. The queue dissolves into negotiation.

The AI assistant displays a justification: \emph{``Triage override applied.''} It offers internal reasons: \emph{``stronger symptoms,'' ``higher risk,''} and a list of contributing factors. It repeats that ``serious illness is prioritized.'' The room keeps arguing: ``So the numbers mean nothing.'' ``Who decides what's serious?'' ``How many overrides are there?'' The reasons do not restore order.

Mae, the senior clerk, steps out and addresses everyone: ``There has been a bus accident in the city. The next two calls will also be emergency walk-ins coming in now. After that, we return to the rule: ticket numbers are the default order; triage emergencies are rare exceptions.'' She points to the ticket dispenser: ``If you don't have a ticket yet, take one here.'' She taps the screen with the banner: \emph{``TRIAGE PRIORITY IN PROGRESS (2). Next ticket: \#10.''} ``This banner is how you'll know an override is happening, and it will always show what number we return to. If your symptoms worsen while you wait, come straight to this desk for reassessment.'' The screen calls \#10, people sit back down, and the queue becomes a queue again.
\end{demonstration}

In the queue demonstration, what fails is not information transfer but how meaning is created in public. Within HCI and AI, research usually starts with framing based on the code-model---senders and receivers encoding/decoding messages over noisy channels---building on an information-theoretic account in which the minds of others remain inaccessible to us. As such, the assistant's banner and its list of factors looks at first like the relevant content: a message to be decoded about why the system ``jumped the line.'' But as \citet{scott2014speaking} stresses, behavioural and textual cues are radically underdetermined: the same observable sequence (repeatedly pulling later numbers up front) could still mean triage, but also a bug, favouritism, or a policy change, and no list of internal features by itself settles which interpretation should organize action. 

In Gricean terms, what is missing is evidence of an intention that is meant to be recognized as such \cite{grice1982meaning, scott2014speaking}; without that ostension, we can say the queue dissolves because the ordering convention---``numbers are the default; exceptions are rare and accountable''---stops being publicly available as a stable, actionable phenomenon. People do not merely believe they are in a queue; they enact it through visible regularities (e.g., picking numbers and waiting) and ordinary repair when those regularities are disrupted (e.g., calling on people skipping the line). When the screen repeatedly jumps ahead without intelligible ostension, patients cannot tell what is happening, what to do next, or how to hold the process to account, so orderly waiting gives way to negotiation and complaint.

Mae's intervention works because it is ostensive and norm-setting rather than post-hoc ``inner reasons.'' She makes the exception legible (``bus accident; next two calls are emergencies''), bounds it (``after that, we return to ticket order''), and anchors it in public affordances (the banner that marks overrides and the machine that issues tickets). This intervention does not reveal the assistant's internal state; it repairs the practice by re-establishing what the system is doing, what people should expect next, and what actions now make sense (sit back down, or come to the desk for reassessment) \cite{dourish2001action, scott2014speaking}. 

Seen through this lens, an ``embodied'' explanation is less a statement that reports inner causes and more an ostensive intervention that makes a situation publicly intelligible and actionable. Because meaning in practice is underdetermined, it must be stabilized through what participants make mutually evident; explanatory force, then, comes from how an intervention reorganizes the shared, intersubjective field of meaning---the shared space of expectations, norms, and action-possibilities that agents continuously sustain and repair together \cite{Gallagher2023, Gallagher2008Intersub}. By treating explanation as shared sense-making, we relocate it to interaction and avoid the ontological obstacles that ``looking inside'' encounters (\Cref{sec:oo}).  

The bigger ontological point is that Mae reinstates the queue \textit{as} a queue by making the norm and its exception publicly legible and actionable: a jointly sustained order of turn-taking---waiting, cutting, exceptions, fairness---stabilized by practices and artifacts that make accountability possible. On this view, explanation is not just ``knowing why'' but---ontologically speaking---what keeps the queue \textit{alive} by making the situation shareable and navigable again: participants regain affordances for orderly waiting, for escalating symptoms, and for holding the process accountable as it unfolds.

\section{Discussion and Future Directions}
\label{sec:discussion}


A counter-argument against the embodied account presented in this paper would be a return to an information-based, internalist account. The AI assistant in \Cref{demo:queue} simply needs more context information: a video feed of the waiting room, access to publicly announced medical emergencies, background on what bus accidents are and what happens when people wait too long, etc. With such information, the assistant could name the bus accident as the culprit for the queue disruption and could utter words similar to Mae. This ``explanation'' could indeed reinstate the queue if we assume that people's attention is sufficiently guided to the assistant's output and does not require the ostensive pointing (``look here'') exhibited by Mae.

We agree that adding the right kinds of context can help---but only insofar as it makes the system more \emph{embodied}, i.e., able to participate in the same public organization of the situation that Mae is involved in. HCI has a long history of designing systems in this spirit: treating meaning as arising in use, building tools around concrete practices, and keeping handles available as resources people can ``act through'' rather than hiding them behind seamless automation \cite{dourish2001action, human-ai-interaction, bohus2019handbook}. Our critique is therefore not aimed at situated, practice-facing systems as such, but at the tendency to treat LLMs as a general-purpose technology and then attempt to ``add explainability'' after the fact via transparency layers or post-hoc XAI. If meaning and understanding are achieved in practice (\Cref{sec:public_explanations}), then explainability requires a commitment to situated participation---being constrained by environments, responsive to downstream consequences, and organized around the concrete activities in which the system is used, rather than presented as a disembodied general-purpose interlocutor.

However, this ``embodiment'' is not what we find in current AI practice, which takes the counter-argument above quite \textit{literally}: context windows with millions of tokens, dataset sizes containing most human text on the internet, and models running on computer clusters consuming the electricity of small cities. We argue that such systems do not become more embodied or situated; rather, they try to model what an embodied agent would say (rather than do) without actually ``having skin in the game''\footnote{Our use of ``skin in the game'' is primarily about \emph{embodied accountability}: acting and bearing the consequences of actions in the same public situation one is helping to organize. ``Mortal computation'' is a related but distinct point here about substrate-bounded intelligence---contrasting portable, exactly replicable ``immortal'' software with computation whose capacities and limits are tied to a particular physical body (finite resources, noise, wear, failure), as emphasized by \citet{hinton2022forwardforward} and developed more formally by \citet{ororbia2024mortalcomputation} via Markov blankets and the free-energy principle.} (being accountable for one's actions, creating meaning through coupling, etc.). Put differently, under the scaling laws of current LLM technology \cite{gpt3}, practice is effectively increasing the model's \textit{epistemic coverage} rather than endowing embodied capacity to enact meaning.

As coverage expands, the cost is epistemic: it becomes increasingly difficult to distinguish situated capability from discursive mimicry. With enough data and compute, LLMs increasingly pass fixed benchmarks---passing theory-of-mind and Turing tests \cite{tom_pass, jones-bergen-2024-gpt}---by producing the right kinds of conversational items. Whether such performance reflects genuine mental-state reasoning or intelligence is contested \cite{lu2025theorymindbenchmarksneed, wang2025rethinkingtheorymindbenchmarks}; our point is prior to that debate. Once you can write the test down as a benchmark, you have already made it the kind of thing scaling can game. Fixed categories turn ``understanding'' into a target distribution; benchmarks then measure convergence to that distribution, not whether the system can keep the phenomenon going \cite{Thomas2022-ei, raji2021ai, koch2021reduced}. On an embodied view, the relevant competence is not ``getting the right answer about the queue,'' but helping the queue remain a queue; and that is precisely what cannot be guaranteed---or even stably recognized---by more epistemic coverage over pre-specified tests.


Likewise, ``explainability'' is often treated as an epistemological mapping---as if progress meant putting the right concepts, values, and norms into the model. In that framing, designers reach for ontological scaffolds---codebooks, feature sets, and ``ground truth'' labels---that delimit what can be known and shown. Such scaffolds can help locally (\Cref{sec_mec-inter}), but they also freeze categories that should remain negotiable as practices and stakes change. Once embedded in systems, these categories can harden into administrative facts that govern people rather than merely describe them; backed by institutional power, the map remakes the territory \cite{alkhatib2021utopia}. In the worst case, labels shape models, models reshape the world, and refreshed labels reinscribe the same ontology---an ``epistemic machinery'' \cite{crawford2021atlas}---while mistaking enacted, shifting meaning for fixed structure.

\begin{sloppypar}
This perspective might clarify a limit of explanation frameworks---and XAI methods (\Cref{sec:xai_llm})---that begin by fixing a set of variables in advance. For example, Pearl-style causal models are extraordinarily powerful once the relevant variables and interventions are agreed upon \cite{pearl1, pearl2}, but they presuppose that the space of causes, effects, and ``what-if'' alternatives is already settled. In many real settings, however, that is precisely what is at issue: what counts as the relevant factor, the appropriate explanation, or even the unit of meaning is often stabilized through practice rather than given upfront (\Cref{sec:cons}) \cite{smith_ooo, Hutto2015-HUTTNO-4}. The queue demonstration makes this re-organization and stabilization vivid and illustrates why pre-packaged ontological scaffolds can fail as explanations for LLM outputs. What restored the queue as a queue was not access to internal, pre-defined factors, but an ostensive, norm-setting repair that re-established a shared space of accountability and action.
\end{sloppypar}

Finally, this ontological lens motivates a  terminological caution in XAI and HCI: token highlights, plausible rationales, and distilled internal stories may be useful, but they should not be treated as explanations by default; they make models explainable only when they work as affordances that make actions publicly available and the situation answerable to what happens next. Accordingly, evaluation should move away from whether users \textit{like} an explanation or find it \textit{plausible}, and toward whether it supports probing, coordination, and repair in situated use. More broadly, we begin not from a demand for internal accounts but from embodied, world-involving practices in which understanding is enacted---often tacit and practical rather than explicit---and we invite HCI to treat as an open question what XAI would look like if it consistently started from those practices, including the possibility that in some cases there is no literal explanation to uncover beyond embodied meaning-making itself.

\section{Conclusion}

In this paper, we asked what it would mean for LLMs to be explainable once we take embodied cognition seriously. Our account of embodied explainability starts from a simple shift: explainability is not a property sitting ``inside'' a model, but an achievement of embodied coupling in practice---people getting a workable grip on what a system is doing as they act with it. \Cref{demo:i-can} used the hammer to show that an object is not just ``matter with properties,'' but something that shows up through skilled use---grasping, swinging, repairing---within shared routines. The wave pushed the same idea to context: it is not a container of facts but a temporally extended interactional achievement. Building on this background, we argue that explainability is an embodied, interactional achievement (\Cref{demo:ontological-problems}): systems become ``explainable'' to users insofar as they provide public affordances that users can appropriate to anticipate, coordinate, and justify action.

Seen through this lens, contemporary ``look inside'' approaches run into ontological obstacles. The first is practical: mechanistic interpretability and other XAI techniques often require heavy abstraction and expert labor, yielding partial, local access that rarely becomes an everyday resource for users at the point of work. The second is deeper: because what counts as ``the thing to be explained'' is constituted in activity---by roles, stakes, bodies, and routines---internal ``explanations'' ultimately miss ``where the action is''. In \Cref{demo:queue}, we thus shift in focus from internal reasons toward how meaning and coordination can be maintained in practice with the help of \textit{public explanations}: they publicly redirect what is salient and what actions are possible so that receivers can participate, test, and repair understanding as the situation unfolds. The upshot is our restrained conclusion: unless LLMs can be made available for dependable coupling---through interactional designs that support anticipation, coordination, contestation, and repair---claims of explainability remain provisional.



\bibliographystyle{ACM-Reference-Format}
\bibliography{sample-base}

\end{document}